\documentclass[aps,prx,twocolumn,groupedaddress,showpacs,showkeys]{revtex4-1}

\usepackage{amsmath}
\usepackage{amssymb}
\usepackage{graphics}
\usepackage{graphicx}
\usepackage{braket}
\usepackage{color}
\usepackage[hidelinks]{hyperref}

\renewcommand{\vec}[1]{\boldsymbol{#1}}

\makeatletter
\newcommand*{\balancecolsandclearpage}{%
  \close@column@grid
  \cleardoublepage
  \twocolumngrid
}

\makeatother

\begin{document}
\title{Electron spin filter and polarizer in a standing light wave}

\author{Sven Ahrens}
\email{ahrens@csrc.ac.cn}

\affiliation{Beijing Computational Science Research Center, Zhongguancun Software Park II, No. 10 West Dongbeiwang Road, Haidian District, Beijing 100193, China}
\date{\today}

\begin{abstract}
We demonstrate the theoretical feasibility of spin-dependent diffraction and spin-polarization of an electron in two counter-propagating, circularly polarized laser beams. The spin-dynamics appears in a two-photon process of the Kapitza-Dirac effect in the Bragg regime. We show the spin-dependence of the diffraction process by comparison of the time-evolution of a spin-up and spin-down electron in a relativistic quantum simulation. We further discuss the spin properties of the scattering by studying an analytically approximated solution of the time-evolution matrix. A classification scheme in terms of unitary or non-unitary propagation matrices is used for establishing a generalized and spin-independent description of the spin properties in the diffraction process.
\end{abstract}





\maketitle


\section{Introduction}

Diffraction of electrons in a standing wave of light as proposed by Kapitza and Dirac \cite{kapitza_dirac_1933_proposal} has been demonstrated at the beginning of the century \cite{Freimund_Batelaan_2001_KDE_first,Freimund_Batelaan_2002_KDE_detection_PRL}, with analogs by diffracting into multiple diffraction orders \cite{Bucksbaum_1988_electron_diffraction_regime} or by using atoms \cite{gould_1986_atoms_diffraction_regime,martin_1988_atoms_bragg_regime}. The Kapitza-Dirac effect has already been studied theoretically, for example in adiabatic switching \cite{fedorov_1974_adiabatic_switching}, by using perturbation theory \cite{gush_1971_electron_scattering,Efremov_1999_classical_and_quantum_KDE,Efremov_2000_wavepacket_theory_KDE}, for spinless particles by using the Klein-Gordon equation \cite{Haroutunian_1975_KDE_analogue,federov_1980_multiphoton_stimulated_compton_scattering}, for the case of a traveling wave in a dielectric medium \cite{hayrapetyan_2015_traveling_wave_KDE} or for a blazed, sawtooth-shaped grating \cite{Vidil_2015_blazed_grating}. See also \cite{batelaan_2007_RMP_KDE} for an overview.

The question, whether the electron spin can be altered in the diffraction process was posed after the observation of the Kapitza-Dirac effect \cite{Batelaan_2003_MSGE,rosenstein_2004_first_KDE_spin_calculation}. Subsequent theoretical considerations confirmed that the electron spin can be manipulated \cite{ahrens_bauke_2012_spin-kde,erhard_bauke_2015_spin,McGregor_Batelaan_2015_two_color_spin,dellweg_awwad_mueller_2016_spin-dynamics_bichromatic_laser_fields} and the dynamical evolution has been identified as a rotation of the electron spin orientation \cite{ahrens_bauke_2013_relativistic_KDE,bauke_ahrens_2014_spin_precession_1,bauke_ahrens_2014_spin_precession_2}. A rotation of the electron spin however does not imply a dependence of the diffraction pattern on the initial spin configuration, nor spin alignment in a certain direction (spin polarization) due to the dynamics.

It is possible to produce polarized electron beams by photoemission \cite{Maruyama_1991_spin_photoemission_InGaAs,Yu_2008_perfect_spin_source}, strong-field ionization \cite{
faisal_2004_spin_intense_field_ionization,barth_2013_spin-polarized_electrons,barth_2014_hole_dynamics_and_spin_currents,
klaiber_2014_relativistic_ionization_spin,klaiber_2014_spin_tunneling,Yakaboylu_2015_ATI_spin,milosevic_2016_spin_polarized_electrons,hartung_2016_spin_strong-field_ionization} and non-linear Compton scattering \cite{panek_2002_laser-induced_compton_scattering}. Also the spin has been investigated in double Compton scattering in a constant crossed field \cite{King_2015_double_compton_scattering_constand_crossed_field} and spin polarization in the magnetic nodes of ultra-intense lasers has been discussed recently \cite{Sorbo_2017_intense_spin_polarization}. Regarding spin-sensitive processes, spin-dependent diffraction has been considered to appear at a phase grating formed by microscopic coils \cite{McGregor_Batelaan_2011_TSGM} or in the near field of a periodic magnetic nano structure \cite{tang_2012_spin_talbot_effect}. Also the possibility of a Stern-Gerlach-like setup for free electrons is discussed in theory \cite{pauli_1932_classical_trajectories,kessler_1976_polarized_electrons,mott_massey_1965_atomic_collisions,Batelaan_1997_electron_stern-gerlach_first,rutherford_grobe_1998_electron_stern-gerlach_comment,Gallup_2001_electron_stern-gerlach_quantum}.

Here we demonstrate that spin-dependent diffraction is possible in a standing light-wave of circularly polarized light for the case of a two-photon interaction. While setups with two interacting photons correspond to the effect considered by Kapitza and Dirac originally \cite{kapitza_dirac_1933_proposal} and have been detected in the experiment \cite{Freimund_Batelaan_2001_KDE_first,Freimund_Batelaan_2002_KDE_detection_PRL} also three-photon scattering has been discussed in bi-chromatic laser fields \cite{smirnova_2004_diffraction_without_grating,dellweg_mueller_2015_bichromatic_KDE} and in particular in the context of spin effects \cite{ahrens_bauke_2012_spin-kde,ahrens_bauke_2013_relativistic_KDE,dellweg_awwad_mueller_2016_spin-dynamics_bichromatic_laser_fields}. In order to show spin-dependent diffraction we explicitly propagate electrons with different initial spin configurations and relate the different outcomes with the initial condition.

Note that at the final stage of our research spin-dependent diffraction has been discussed in the context of a three-photon interaction in the Kapitza-Dirac effect \cite{dellweg_mueller_2016_interferometric_spin-polarizer,dellweg_mueller_extended_KDE_calculations}. The former setup \cite{dellweg_mueller_2016_interferometric_spin-polarizer} makes use of an interferometric setup with linear polarized laser beams, which combines three-photon and two-photon Kapitza-Dirac scattering, while the latter setup \cite{dellweg_mueller_extended_KDE_calculations} solely  considers three-photon scattering in laser fields with a general polarization description. The effective three-photon interaction is realized by employing a bi-chromatic standing light wave as external field. In contrast to that, we are investigating Kapitza-Dirac scattering in a mono-chromatic, standing light wave of circular polarization, in which the electron is undergoing an effective two-photon interaction.

Our article is organized as follows: In section \ref{eq:laser_field} we describe the laser field and in section \ref{eq:relativistic_equations_of_motion} we introduce the notion of relativistic quantum dynamics in momentum space. We present a full simulation of the equations of motion in \ref{eq:spin-properties_simulation} with a Gaussian-shaped, temporal interaction of the electron with the laser field, and we point out spin-dependent diffraction and spin-polarization effects. In section \ref{sec:approximative_solution} we adapt an approximate solution of the quantum dynamics \cite{erhard_bauke_2015_spin} for a laser field which propagates along the $z$-axis. Based on this solution we give an intuitive explanation for the origin of the described spin-dependent diffraction in Sec. \ref{sec:explanation_of_process}. Section \ref{sec:properties_of_spin_dynamics} discusses the spin properties of the analytic solution by starting with general considerations on the degrees of freedoms of the $2\times 2$ submatrix which is responsible for the propagation of the electron spin. We further approximate the solution for different time-scales, ie. at instant times in section \ref{sec:instant_rabi_oscillations} and after an eighth of the period $2 \pi/\Omega_S$ in section \ref{sec:eigth_period_approximation}, where $\Omega_S$ is a characteristic frequency of spin effects. For the latter case we investigate the spin-dependent diffraction in section \ref{sec:spin-dependent_diffraction} and spin-polarization of the electron in section \ref{sec:electron_spin-polarization}  by comparing the dynamics with a more accurate analytic solution in appendix \ref{sec:more_acurate_solution} and the numerical simulation of the quantum dynamics. In section \ref{eq:full_spin_filter} we discuss the extremal cases of the spin-dependent diffraction. We conclude in section \ref{sec:conclusions} with outlining the implications of the discussed spin-dynamics.

\section{Simulation of relativistic quantum dynamics\label{sec:simulation}}

For the description of the process we solve the quantum dynamics of the single particle Dirac equation
\begin{equation}
 i \hbar \dot \Psi(\vec x,t) = \left[ c \left(- i \hbar \vec \nabla - \frac{q}{c} \vec A(\vec x,t) \right)\cdot \vec \alpha + m c^2 \beta \right] \Psi(\vec x,t)\,,\label{eq:dirac-equation}
\end{equation}
in momentum space by making use of a plane wave expansion of the wave function \cite{ahrens_bauke_2012_spin-kde,ahrens_bauke_2013_relativistic_KDE,bauke_ahrens_2014_spin_precession_1,bauke_ahrens_2014_spin_precession_2}. The constants in Eq. \eqref{eq:dirac-equation} are the reduced Planck constant $\hbar$, the electron rest mass $m$ and the vacuum speed of light $c$. The $\alpha_i$ of the vector $\vec \alpha$ and $\beta$ are the Dirac matrices. In this article we use a dot above a time-dependent variable to denote its time derivative, for example $\partial \Psi(\vec x,t)/\partial t = \dot \Psi(\vec x,t)$.

\subsection{The external electro-magnetic field\label{eq:laser_field}}

We describe the vector field of two counter-propagating, circularly polarized laser beams by
\begin{equation}
 \vec A(\vec x, t) = 2 A w(t) \cos(k z) \left[ - \sin(\omega t) \vec e_x + \cos(\omega t) \vec e_y \right] \,,\label{eq:vector_potential}
\end{equation}
with wavenumber $k$ and frequency $\omega=ck$. Eq. \eqref{eq:vector_potential} is a solution of the Maxwell equations, provided that the envelope function $w(t)$ was constant in time. We model the temporal interaction of the electron with the laser beam in our numerical simulation by using the envelope function
\begin{equation}
 w(t) = \sin^2 \left(\frac{\pi t}{\tau}\right)\,,\label{eq:beam-envelope}
\end{equation}
in accordance with earlier studies \cite{ahrens_bauke_2012_spin-kde,ahrens_bauke_2013_relativistic_KDE,bauke_ahrens_2014_spin_precession_1,bauke_ahrens_2014_spin_precession_2,dellweg_mueller_2015_bichromatic_KDE,dellweg_mueller_2015_KDE_pulse_shape,erhard_bauke_2015_spin}. The parameter $\tau$ is the time period of the interaction and the simulation is evolved in the period between 0 and $\tau$, ie. $t\in[0,\tau]$.

We point out that dynamics in standing light waves have been studied by the investigation of classical trajectories \cite{Kirk_2016_radiative_trapping}, by solving the Klein-Gordon equation \cite{King_2016_scalar_quantum_dynamics,Varro_2014_Klein-Gordon_solutions} or studying non-linear Compton scattering \cite{Raicher_2016_nonlinear_compton_scattering}.


\subsection{Equations of motion in momentum space\label{eq:relativistic_equations_of_motion}}

For the description of the relativistic wave function, we introduce the bi-spinors
\begin{subequations}
\begin{align}
 u_n^{+,\alpha} &= \sqrt{\frac{\mathcal{E}_n + m c^2}{2 \mathcal{E}_n}}
 \begin{pmatrix}
  \chi^\alpha \\ \frac{n c k \hbar \sigma_z}{\mathcal{E}_n + m c^2} \chi^\alpha
 \end{pmatrix}\label{eq:bi-spinors_positive}\\
 u_n^{-,\alpha} &= \sqrt{\frac{\mathcal{E}_n + m c^2}{2 \mathcal{E}_n}}
 \begin{pmatrix}
  - \frac{n c k \hbar \sigma_z}{\mathcal{E}_n + m c^2} \chi^\alpha \\ \chi^\alpha
 \end{pmatrix}\,,\label{eq:bi-spinors_negative}
\end{align}\label{eq:bi-spinors}\\
\end{subequations}
and the solutions of the free Dirac equation \cite{ahrens_bauke_2012_spin-kde,ahrens_bauke_2013_relativistic_KDE}
\begin{equation}
 \psi_n^{\gamma,\alpha}(\vec x) = \sqrt{\frac{k}{2 \pi}} u_n^{\gamma,\alpha} e^{i n k z}\,. \label{eq:relativistic_momentum_eigenstates}
\end{equation}
Here, $\mathcal{E}_n$ is the relativistic energy momentum relation
\begin{equation}
 \mathcal{E}_n = \sqrt{(m c^2)^2 + (n c k \hbar)^2}
\end{equation}
and $\sigma_x$, $\sigma_y$ and $\sigma_z$ are the three Pauli matrices. The two component objects $\chi^{\uparrow} = (1,0)^T$, $\chi^{\downarrow} = (0,1)^T$ form a basis in the spinor space of the Pauli equation. The functions \eqref{eq:relativistic_momentum_eigenstates} are simultaneous eigenfunctions of the momentum operator, the free Dirac Hamiltonian and of the Foldy-Wouthuysen spin operator \cite{Foldy_Wouthuysen_1950_FW-transformation}. Correspondingly, the quantum numbers of the eigenfunctions denote the momentum $n k \hbar$, the sign of the eigenenergy $\gamma \in \{+,-\}$ and the spin $\sigma \in \{ \uparrow, \downarrow \}$.

The relativistic wavefunction of the electron 
\begin{equation}
 \Psi(\vec x,t) = \sum_{n \in \mathbb{N}, \atop \alpha\in\{\uparrow,\downarrow\}} c_n^\alpha(t) \psi_n^{+,\alpha} + d_n^\alpha(t) \psi_n^{-,\alpha} \label{eq:relativistic_wave_function}
\end{equation}
is expanded in terms of the described eigenfunctions, where the time-evolution of the expansion coefficients $c_n^\alpha(t)$ and $d_n^\alpha(t)$ is obtained by projecting the Dirac equation \eqref{eq:dirac-equation} at the plane waves \eqref{eq:relativistic_momentum_eigenstates}. We obtain a system of differential equations
\begin{subequations}%
\begin{multline}%
 i \hbar \dot c_n^\alpha(t) = \phantom{-}\mathcal{E}_n c_n^\alpha(t) + \sum_{n' \in \mathbb{N}, \atop \beta\in\{\uparrow,\downarrow\}} \Bigg[ V_{n,n'}^{+,\alpha;+,\beta}(t) c_{n'}^\beta(t) \\
 + V_{n,n'}^{+,\alpha;-,\beta}(t) d_{n'}^\beta(t) \Bigg]\,,
\end{multline}
\begin{multline}
 i \hbar \dot d_n^\alpha(t) = -\mathcal{E}_n d_n^\alpha(t) + \sum_{n' \in \mathbb{N}, \atop \beta\in\{\uparrow,\downarrow\}} \Bigg[ V_{n,n'}^{-,\alpha;+,\beta}(t) c_{n'}^\beta(t) \\
 + V_{n,n'}^{-,\alpha;-,\beta}(t) d_{n'}^\beta(t) \Bigg]\,,
\end{multline}\label{eq:dirac_equation_momentum_space}%
\end{subequations}%
with the interaction term
\begin{multline}
 V_{n,n'}^{\gamma,\rho;\gamma',\rho'}(t) = \frac{q}{c} A \, w(t) \left( \delta_{n,n'-1} + \delta_{n,n'+1} \right)\\
 \cdot u_n^{\gamma,\rho\dagger} \left[\alpha_1 \sin(\omega t) - \alpha_2 \cos(\omega t)\right] u_n^{\gamma',\rho'} \,.\label{eq:momentum_space_dirac_equation}
\end{multline}

Note, that in the numerical simulation, the amplitude $|c_n^\alpha(t)|$ and $|d_n^\alpha(t)|$ is dropping exponentially for large $|n|$. Therefore we truncate the system of differential equations of the expansion coefficients and set them to zero for $|n|>10$.

%
\begin{figure*}[!ht]%
  \includegraphics[width=0.9\textwidth]{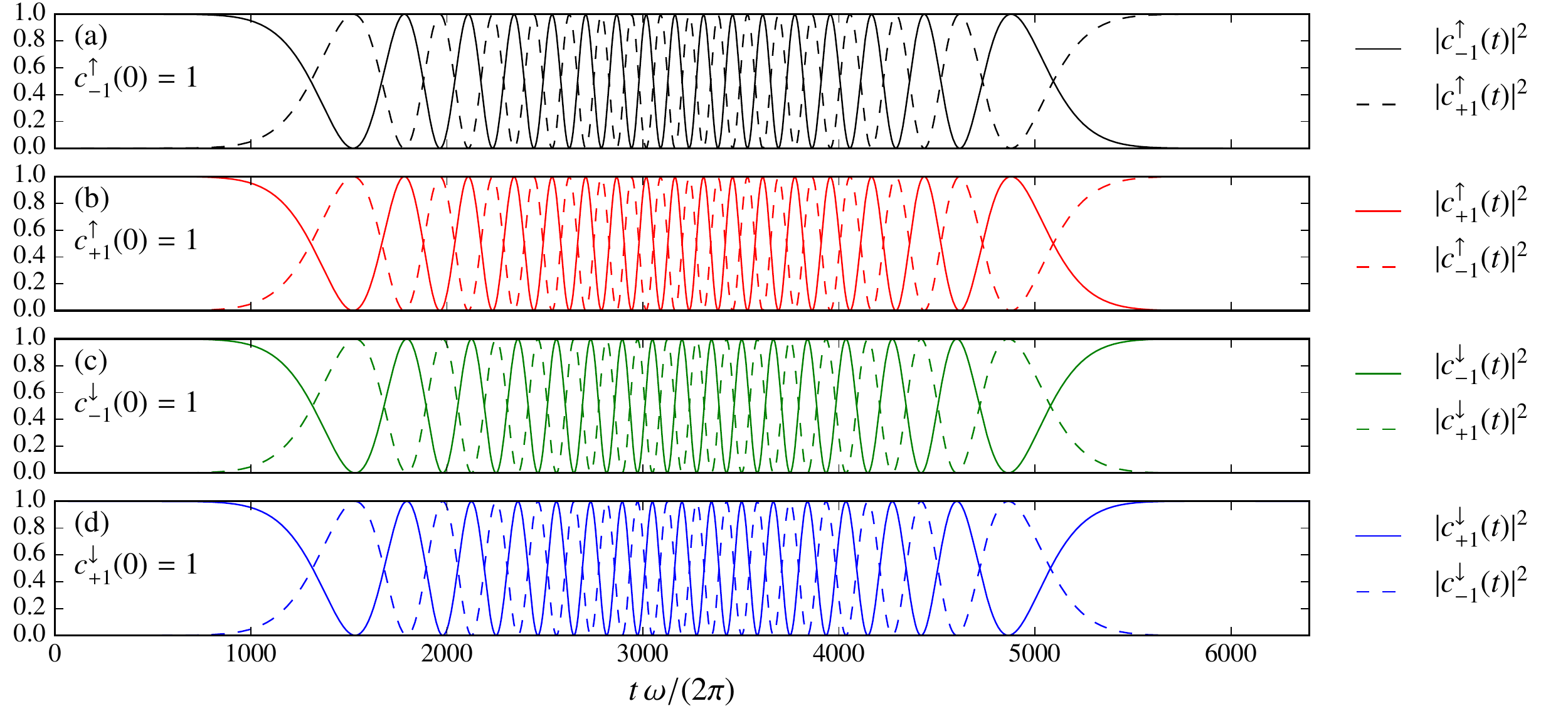}%
  \caption{\label{fig:simulation}
(Color online) Simulated quantum dynamics. Shown are the absolute squares of the non-vanishing expansion coefficients of the wavefunction \eqref{eq:relativistic_wave_function} over the time $t$. For the initial condition we set the the expansion coefficient $c_{-1}^\uparrow(0)$, $c_{+1}^\uparrow(0)$, $c_{-1}^\downarrow(0)$ and $c_{+1}^\downarrow(0)$ to one in the four subfigures (a), (b), (c) and (d) respectively, where all other expansion coefficients $c_n^\sigma(0)$ and $d_n^\sigma(0)$ are set to zero at time $t=0$. One can see, that electrons with spin-up polarization are reversing their momentum, while the electrons with spin-down polarization are not changing their momentum. This means that spin-dependent diffraction is taking place. The laser peak intensity is $1.12 \times 10^{22}\,\textrm{W}/\textrm{cm}^2$, with the wavelength $\lambda=0.159\,\textrm{nm}$ in our simulation, in accordance with the parameters used in \cite{erhard_bauke_2015_spin}.  
}%
\end{figure*}%
%
%
\begin{figure}%
  \includegraphics[width=0.25\textwidth]{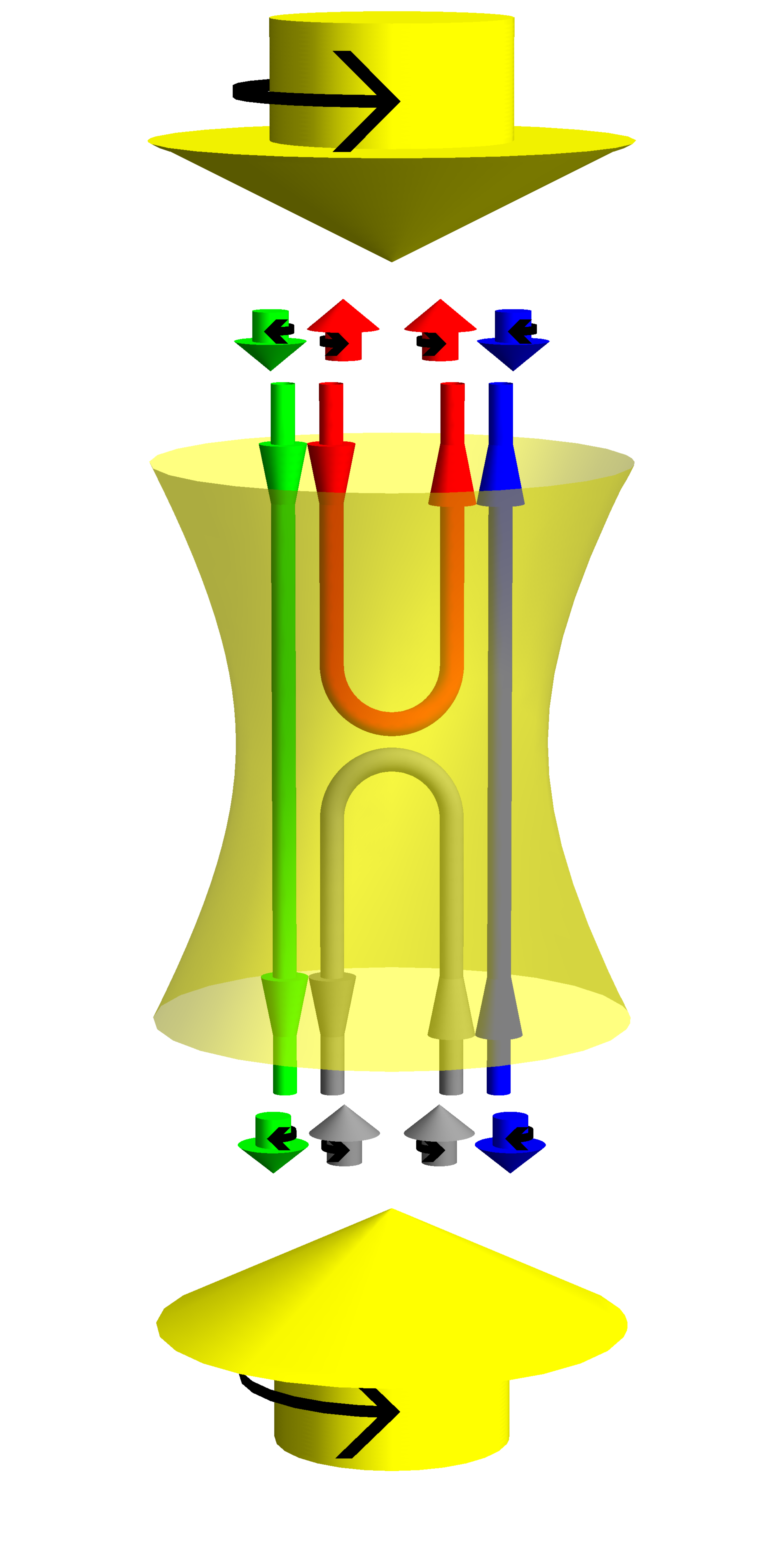}%
  \caption{\label{fig:experimental_setup}
(Color online) Interacting setup of laser and electron. Sketched in yellow arrows are two counter-propagating, co-rotating, circularly polarized laser beams from top and bottom and a corresponding yellow hyperboloid, which illustrates the resulting standing light wave. Depicted on the upper left and lower right are incoming spin-up and spin-down electrons, while on the lower left and the upper right are the outgoing electrons. The connecting lines of electrons indicate that momenta of spin-up electrons are reversed, while momenta of spin-down electrons remain unchanged by interaction with the laser field. This property is consistent with the simulation in Fig. \ref{fig:simulation}. The diagram also coincides with the analytic expression \eqref{eq:specific_R_T_matrices} at time parameter $\eta=8\cdot2\pi$, where $\eta$ is defined in Eq. \eqref{eq:def_eta} (see also section \ref{eq:full_spin_filter}).
}%
\end{figure}%
%

\subsection{The numerical simulation and its spin properties\label{eq:spin-properties_simulation}}

We demonstrate the possibility of filtering and polarizing the electron spin by a simulation of the eigensolutions's expansion coefficients $c_n^\sigma(t)$ and $d_n^\sigma(t)$, shown in Fig. \ref{fig:simulation}. The electron spin is initially pointing upwards in the subfigures \ref{fig:simulation} (a) and \ref{fig:simulation} (b) and it initially points downwards in the Figures \ref{fig:simulation} (c) and \ref{fig:simulation} (d). Likewise, the initial electron is moving in the $-z$ direction with momentum $\hbar k = 7.8\,\textrm{keV}/c$ in the subfigures \ref{fig:simulation} (a) and \ref{fig:simulation} (c) and it is moving with momentum $\hbar k$ in the $z$ direction in the subfigures \ref{fig:simulation} (b) and \ref{fig:simulation} (d). One can see that spin-up electrons change their initial occupation probabilities $|c_{\pm 1}(0)|^2=1$ and $|c_{\mp 1}(0)|^2=0$ to the final probabilities $|c_{\pm 1}(\tau)|^2=0$ and $|c_{\mp 1}(\tau)|^2=1$, where
\begin{equation}
 |c_n(t)|^2 = |c_n^\uparrow(t)|^2 + |c_n^\downarrow(t)|^2 \label{eq:general_diffraction_probability}
\end{equation}
is the probability of finding the electron with momentum $n \hbar k$ in $z$-direction at time $t$. In contrast, spin-down electrons are not exchanging their occupation probabilities between the momenta $\hbar k$ and $-\hbar k$. This spin-dependent diffraction behavior implies that it is possible to separate electrons according to their spin-state in the 2-photon Kapitza-Dirac effect with circularly polarized light.

Also, if the electron is initially moving upwards with momentum $\hbar k$, as in the subfigures \ref{fig:simulation} (b) and \ref{fig:simulation} (d) then the final electron will move downwards with electron spin pointing up or move upwards with electron spin pointing down. Similarly, an initially downwards moving electron as in the subfigures \ref{fig:simulation} (a) and \ref{fig:simulation} (c) will finally move upwards with spin pointing up or move downwards with spin pointing down. This means that the electron spin is polarized independent of its initial spin configuration.

The described spin dynamics are sketched in Fig. \ref{fig:experimental_setup} for illustration.

\section{Approximate description\label{sec:approximative_solution}}

A simplified picture of the spin dynamics in the considered system can be given by an approximate analytic solution of the relativistic quantum dynamics, which is described by Erhard and Bauke \cite{erhard_bauke_2015_spin}. The method makes use  of a Foldy-Wouthuysen transformation of the Dirac equation for obtaining a non-relativistic approximation \cite{Foldy_Wouthuysen_1950_FW-transformation,froehlich_studer_1993_FW-transformation}. The non-negligible contributions of the transformation can be written in terms of the Pauli equation plus a relativistic correction term
\begin{multline}
   i \hbar \dot \Psi(\vec x,t) = \Bigg\{ \frac{1}{2 m}\left[-i \hbar \vec \nabla - \frac{q}{c} \vec A(\vec x,t)\right]^2 - \frac{q \hbar}{2 m c} \vec \sigma \cdot \vec B(\vec x,t) \\
   + \frac{q^2 \hbar}{4 m^2 c^3} \vec \sigma \cdot \left[ \vec E(\vec x,t) \times \vec A(\vec x,t) \right]\Bigg\}\Psi(\vec x,t) \,,\label{eq:relativistic_pauli_equation}
\end{multline}
where a constant $m c^2$ contribution is neglected due to the possible elimination by choosing a suitable gauge. The electric and magnetic fields are related to the vector potential \eqref{eq:vector_potential} by
\begin{align}
 \vec E(\vec x,t) &= - \frac{1}{c}\frac{\partial \vec A(\vec x,t)}{\partial t} \quad \textrm{and} \\ \vec B(\vec x,t) &= \nabla \times \vec A(\vec x,t)\,.
\end{align}
The approximation from Erhard and Bauke assumes a non-varying field amplitude $w(t)=1$, for which the electric and magnetic fields evaluate to
\begin{subequations}%
\begin{align}%
 \vec E(\vec x, t) &= 2 A \, k \cos(k z) \left[ \cos(\omega t) \vec e_x + \sin(\omega t) \vec e_y \right]\,, \label{eq:electric_field}\\
 \vec B(\vec x, t) &= 2 A \, k \sin(k z) \left[ \cos(\omega t) \vec e_x + \sin(\omega t) \vec e_y \right]\,.\label{eq:magnetic_field}
\end{align}\label{eq:electro-magnetic_field}%
\end{subequations}%
In another step \cite{erhard_bauke_2015_spin}, Equation \eqref{eq:relativistic_pauli_equation} can be solved with the Magnus expansion \cite{magnus_1954_magnus_expansion,blanes_casas_2009_magnus_expansion}, where terms which are negligibly small and which are not growing linearly in time are neglected. The relevant terms of the calculation appear in an exponential representation of the time-evolution and correspond to the wave equation
\begin{multline}
 i \hbar \dot \Psi(\vec x,t) = \Bigg[ \frac{1}{2 m}\left(-i \hbar \vec \nabla \right)^2 + \frac{2 q^2 {A}^2}{m c^2} \cos^2(k z) \\
 - \frac{q^2 {A}^2 \hbar k}{m^2 c^3} \left[ \sin^2(k z) - \cos^2(k z) \right] \sigma_z \Bigg] \Psi(\vec x,t)\,.\label{eq:magnus_relativistic_pauli_equation}
\end{multline}
One can exchange the relativistic wave function \eqref{eq:relativistic_wave_function} of the Dirac equation by the two-component wave function
\begin{equation}
 \Psi(\vec x,t) = \sum_{n} c_n(t) e^{i n k z} \label{eq:pauli_wave_function}\,,
\end{equation}
with the two-component structure
\begin{equation}
c_n(t) =
\begin{pmatrix}
 c_n^\uparrow(t) \\ c_n^\downarrow(t)
\end{pmatrix}
\end{equation}
of expansion coefficients for the case of the Foldy-Wouthuysen transformed wave equation \eqref{eq:relativistic_pauli_equation}. Inserting the wave function \eqref{eq:pauli_wave_function} into the relativistic Pauli equation \eqref{eq:magnus_relativistic_pauli_equation} and projecting on the plane-wave eigenfunctions $\chi^\sigma e^{i n k z}$ yields
\begin{multline}
 i \hbar \dot c_n(t) = \frac{n^2 \hbar^2 k^2}{2 m} c_n(t) \\
 + \frac{q^2 {A}^2}{2 m c^2} \left[ c_{n-2}(t) + 2 c_n(t) + c_{n+2}(t) \right] \\
 + \frac{q^2 {A}^2 \hbar k}{2 m^2 c^3} \sigma_z \left[ c_{n-2}(t) + c_{n+2}(t) \right] \,.\label{eq:momentum_space_pauli_equation}
\end{multline}
We restrict the system of differential equations to the electron momenta $-\hbar k$ and $\hbar k$, according to a similar description of the Kapitza-Dirac effect in the so-called Bragg regime, which is discussed by Batelaan  \cite{batelaan_2000_KDE_first,Freimund_Batelaan_2002_KDE_detection_PRL,batelaan_2007_RMP_KDE}, resulting in
\begin{equation}
 i \dot c_{\pm 1}(t) = (\Omega_k + 2 \Omega_R) c_{\pm 1}(t) + (\Omega_R \mathbf{1} + \Omega_S \sigma_z) c_{\mp 1}(t)\,.\label{eq:direct_simplified_pauli_equation}
\end{equation}
Analogous to \cite{erhard_bauke_2015_spin}, we have introduced the abbreviations for the kinetic energy
\begin{equation}
 \Omega_k = \frac{ \hbar k^2}{2 m}
\end{equation}
the Rabi frequency of the Kapitza-Dirac effect
\begin{equation}
 \Omega_R = \frac{q^2 {A}^2}{2 m c^2 \hbar}\,,
\end{equation}
and the frequency of spin-dependent effects
\begin{equation}
 \Omega_S = \Omega_R \frac{\hbar k}{m c}\,.
\end{equation}
here. By choice of a suitable gauge, the term proportional to $(\Omega_k + 2 \Omega_R)$ can be removed, yielding
\begin{equation}
 i \dot c_{\pm 1}(t) = (\Omega_R \mathbf{1} + \Omega_S \sigma_z) c_{\mp 1}(t)\,.\label{eq:simplified_pauli_equation}
\end{equation}
By introducing the notion
\begin{equation}
 c_{\pm1}(t) = T(t) c_{\pm1}(0) + R(t) c_{\mp1}(0) \label{eq:time-evolution-notion}
\end{equation}
for the time evolution one can write the solution of the differential equation Eq. \eqref{eq:simplified_pauli_equation} as
\begin{subequations}
\begin{align}
T(t)&=
\begin{pmatrix} \cos[(\Omega_R + \Omega_S)t] & 0 \\
0 & \cos[(\Omega_R - \Omega_S)t]
\end{pmatrix}\label{eq:propagator_A}\\
R(t)&= -i
\begin{pmatrix} \sin[(\Omega_R + \Omega_S)t] & 0 \\
0 & \sin[(\Omega_R - \Omega_S)t]
\end{pmatrix}\,.\label{eq:propagator_B}
\end{align}\label{eq:propagator_submatricies}%
\end{subequations}
Here, the matrices $R(t)$ and $T(t)$ describe spin-dependent initial- to final quantum state scattering with and without reversion of momentum, corresponding to reflection and transmission.

Note, that it is possible to solve the system of equations \eqref{eq:simplified_pauli_equation} also by accounting for the momenta $\pm 3 \hbar k$, instead of $\pm 1 \hbar k$ only \cite{erhard_bauke_2015_spin}, as explained in appendix \ref{sec:more_acurate_solution}. One obtains corrections to the solution \eqref{eq:propagator_submatricies} which are negligibly small for the parameters in Fig. \ref{fig:simulation}.

\section{Explanation of the physical process\label{sec:explanation_of_process}}

One can intuitively understand the spin-dependent quantum dynamics by looking at Eq. \eqref{eq:simplified_pauli_equation} and Eq. \eqref{eq:propagator_submatricies}. The spin-up and spin-down components in Eq. \eqref{eq:simplified_pauli_equation} are not mixing and consequently the spin propagation matrices \eqref{eq:propagator_submatricies} are diagonal. However, one can see that the Rabi-flopping frequency $\Omega_R$ of the spin-up component is enhanced by the small value $\Omega_S$, while the spin-down component is decreased by $\Omega_S$. Thus, after elapsing several Rabi cycles there is a time at which the Rabi cycle of a spin-up electron is complete, while the Rabi cycle of a spin-down electron is not, or vice versa. This property can also be observed in the numerical simulation in Fig. \ref{fig:simulation}, in which the spin-up electron oscillates through 16.5 cycles, while the spin-down electron only evolves 16 cycles. As a result only a spin-up electron will be in the diffracted state, while a spin-down electron remains in its initial state, corresponding to spin-dependent diffraction. In this sense, the spin effect presented here is caused by stronger (weaker) interaction of electrons with spin co-aligned (counter-aligned) to the spin-density of the laser beam, respectively.


\section{Properties of spin dynamics\label{sec:properties_of_spin_dynamics}}

The matrices $T(t)$ and $R(t)$ in Eq. \eqref{eq:propagator_submatricies} can be represented in the form \cite{ahrens_bauke_2013_relativistic_KDE}
\begin{equation}
 U_s(t)=\sqrt{P} e^{i \chi} \left[ \cos \left(\frac{\xi}{2}\right) \mathbf{1} - i \sin\left(\frac{\xi}{2}\right) \vec n \cdot \vec \sigma \right]\,.\label{eq:su2-representation}
\end{equation}
For the case of real parameters $P$, $\chi$, $\xi$ and $\vec n$ (with $\vec n$ being a unit vector) the term in the square brackets is an $\textrm{SU}(2)$ rotation which rotates the electron spin. In this case $P$ has the meaning of a spin-independent diffraction probability. In general the unit vector $\vec n$ can also be complex-valued, such that the representation \eqref{eq:su2-representation} is parameterized by 8 independent parameters, corresponding to the 8 degrees of freedom of a complex valued $2\times 2$ matrix \cite{ahrens_2012_phdthesis_KDE}. In this more general case the term in the square bracket is no longer an $\mathcal{SU}(2)$ rotation and the parameter $P$ looses it's property as spin-independent diffraction probability, because different spin-configurations will be diffracted with different probability.

We want to demonstrate this property for the time-propagation \eqref{eq:propagator_submatricies}. The solution \eqref{eq:propagator_submatricies} consists of a fast dynamical part of scale $\Omega_R$ and a slow dynamical part of scale $\Omega_S$ which one can see from the expansion of the trigonometric functions
\begin{align}
 \cos\left[(\Omega_R \pm \Omega_S) t\right] &= \cos \Omega_R t \cos \Omega_S t \mp \sin \Omega_R t \sin \Omega_S t \nonumber\\
 \sin\left[(\Omega_R \pm \Omega_S) t\right] &= \sin \Omega_R t \cos \Omega_S t \pm \cos \Omega_R t \sin \Omega_S t\,.\label{eq:trigonometric_expansion}
\end{align}

\subsection{Instant Rabi-oscillations\label{sec:instant_rabi_oscillations}}
We note that $\Omega_S$ is smaller than $\Omega_R$ by a factor of $\hbar k/(mc)$, ie. 
$\Omega_R = 65.4 \, \Omega_S$ for the parameters used in the simulation of Fig. \ref{fig:simulation}. Therefore, for times $t \ll 2\pi/\Omega_S$, the $\sin(\Omega_S t)$ term can be neglected and the $\cos(\Omega_S t)$ term can be set to one in Eq. \eqref{eq:trigonometric_expansion}, resulting in 
\begin{subequations}
\begin{equation}
T(t) \approx
 \begin{pmatrix}
  \cos \Omega_R t  & 0 \\
  0 & \cos \Omega_R t
 \end{pmatrix}
\end{equation}
and
\begin{equation}
R(t) \approx -i
 \begin{pmatrix}
  \sin \Omega_R t & 0 \\
  0 & \sin \Omega_R t
 \end{pmatrix}\,.
\end{equation}\label{eq:instant_rabi_oscillations}
\end{subequations}
The matrix \eqref{eq:su2-representation} can approximate $T(t)$ and $R(t)$ by setting
\begin{subequations}
\begin{align}
 \xi&=0,\ P=\cos^2 \Omega_R t \qquad \textrm{for } T(t) \textrm{ and}\\
 \xi&=0,\ P=\sin^2 \Omega_R t \qquad \textrm{for } R(t)\,,
\end{align}
\end{subequations}
corresponding to spin-independent Rabi-oscillations in the Kapitza-Dirac effect \cite{batelaan_2007_RMP_KDE}.

\subsection{After an eighth of period $2 \pi/\Omega_S$\label{sec:eigth_period_approximation}}

The dynamics change, when the product $\Omega_S t$ reaches fractions of the period $2 \pi$. An interesting value is an eighth of the period $2 \pi/\Omega_S$, which also corresponds to the final quantum state configuration of the simulation in Fig. \ref{fig:simulation}, as explained in section \ref{eq:full_spin_filter}. For further investigations we introduce the shifted time
\begin{equation}
 t' = t - \frac{\pi}{4 \Omega_S}\,.\label{eq:shifted_time_quarter}
\end{equation}
The functions $\cos \Omega_S t$ and $\sin \Omega_S t$ evaluate to $1/\sqrt{2}$ for $|t'| \ll 2 \pi/\Omega_S$ in the trigonometric expansion \eqref{eq:trigonometric_expansion}, resulting in the non-vanishing matrix elements
\begin{align}
 T_{11}(t) &\approx [ \cos(\Omega_R t' + \varphi_0) - \sin(\Omega_R t' + \varphi_0) ]/\sqrt{2}\nonumber\\
 T_{22}(t) &\approx [ \cos(\Omega_R t' + \varphi_0) + \sin(\Omega_R t' + \varphi_0) ]/\sqrt{2}\nonumber\\
 R_{11}(t) &\approx -i [ \sin(\Omega_R t' + \varphi_0) + \cos(\Omega_R t' + \varphi_0) ]/\sqrt{2}\nonumber\\
 R_{22}(t) &\approx -i [ \sin(\Omega_R t' + \varphi_0) - \cos(\Omega_R t' + \varphi_0) ]/\sqrt{2}\nonumber\\
 & \label{eq:propagation_quarter_approximation}
\end{align}
where we introduce the abbreviation
\begin{equation}
 \varphi_0 = \frac{\pi}{4}\frac{\Omega_R}{\Omega_S} \,.
\end{equation}
The matrices \eqref{eq:propagation_quarter_approximation} can be expressed in terms of the spin approximation $U_s(t)$ in Eq. \eqref{eq:su2-representation}, by using the parameters 
\begin{subequations}%
\begin{align}%
 P &= 1/2\,,\quad \chi=0 \,,\quad \vec n = (0,0,-i)^T\,,\label{eq:su-rotation_adaption_first_line}\\
 \xi &= 2 \Omega_R t' + 2 \varphi_0 \quad \textrm{for } T(t)\label{eq:su-rotation_adaption_second_line}
\end{align}
and
\begin{align}%
 P &= 1/2\,,\quad \chi=3 \pi/2 \,,\quad \vec n = (0,0,-i)^T\,,\label{eq:su-rotation_adaption_first_line2}\\
 \xi &= 2 \Omega_R t' + 2 \varphi_0 - \pi  \quad \textrm{for } R(t)\,.\label{eq:su-rotation_adaption_second_line2}
\end{align}\label{eq:su-rotation_adaption}
\end{subequations}%
The matrix in the square brackets in Eq. \eqref{eq:su2-representation} is no longer an $\mathcal{SU}(2)$ matrix, for the set of parameters in Eq. \eqref{eq:su-rotation_adaption}, because the vector $\vec n$ is imaginary-valued now. It implies that the diffraction probability depends on the electron's spin-configuration, where spin-dependent diffraction is characterized as a non-unitary propagation matrix of the electron spin.

For investigating the spin properties of the diffraction process, it is suitable to shift the time \eqref{eq:shifted_time_quarter} further by the small value $\pi/(4 \Omega_R)$. Therefore, we introduce the shifted time
\begin{equation}
 t^{\prime\prime} = t - \frac{\pi}{4\Omega_R} - \frac{\pi}{4\Omega_S}\,.\label{eq:shifted_time_quarter_plus}
\end{equation}
Inserting this shifted time in the arguments of the trigonometric functions \eqref{eq:propagator_submatricies} yields
\begin{subequations}
\begin{align}
 \Omega_R t + \Omega_S t &= \Omega_R t^{\prime\prime} + \varphi_0 + \Omega_S \left(t^{\prime\prime} + \frac{\pi}{4 \Omega_R} \right) + \frac{\pi}{2} \label{eq:phase-term_sum}\\
 \Omega_R t - \Omega_S t &= \Omega_R t^{\prime\prime} + \varphi_0 - \Omega_S \left(t^{\prime\prime} + \frac{\pi}{4 \Omega_R} \right)\,.\label{eq:phase-term_diff}
\end{align}\label{eq:phase-term}%
\end{subequations}
The term $\Omega_S[t^{\prime\prime} + \pi/(4 \Omega_R)]$ is negligibly small for times $|t^{\prime\prime}|\ll 2 \pi/\Omega_S$ and can be omitted, such that the matrices $T(t)$ and $R(t)$ in \eqref{eq:propagator_submatricies} can be written as 
\begin{equation}%
T(t)\approx
\begin{pmatrix} -\sin\eta & 0 \\
0 & \cos \eta
\end{pmatrix},\quad
R(t)\approx -i
\begin{pmatrix} \cos \eta & 0 \\
0 & \sin \eta
\end{pmatrix},\label{eq:spin-rabi-propagator-submatrices}%
\end{equation}
where the argument of the trigonometric functions is abbreviated with
\begin{equation}
 \eta=\Omega_R t^{\prime\prime} + \varphi_0\,.\label{eq:def_eta}
\end{equation}
The parameterization of $U_s(t)$ for Eq. \eqref{eq:spin-rabi-propagator-submatrices} is analogous to Eq. \eqref{eq:su-rotation_adaption}, where only $\xi$ needs to be exchanged and be expressed in terms of $t^{\prime\prime}$, yielding
\begin{subequations}
\begin{equation}
 \xi = 2 \Omega_R t^{\prime\prime} + 2 \varphi_0 + \pi/2\,,\label{eq:shifted_xi_parameter}
\end{equation}
instead of Eq. \eqref{eq:su-rotation_adaption_second_line} for $R(t)$ and
\begin{equation}
 \xi = 2 \Omega_R t^{\prime\prime} + 2 \varphi_0 - \pi/2\label{eq:shifted_xi_parameter2}
\end{equation}
\end{subequations}
instead of Eq. \eqref{eq:su-rotation_adaption_second_line2} for $T(t)$.

\subsection{Spin-dependent diffraction\label{sec:spin-dependent_diffraction}}

\begin{figure}%
\vspace{0.7 cm}%
  \includegraphics[width=0.35\textwidth]{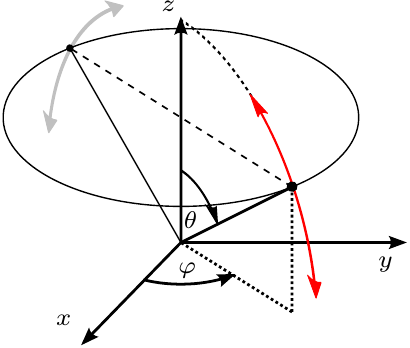}%
  \caption{\label{fig:bloch_vector}%
(Color online) Bloch vector in our setup. Shown is the expectation value of the quantum state \eqref{eq:bloch_state} with respect to the vector $\vec \sigma$ of Pauli matrices, being the Bloch vector, which points in the direction $(\theta,\varphi)$ in spherical coordinates. The change of the vector's $z$-component due to the laser-electron interaction in Fig. \ref{fig:Bloch_z_component} is indicated by the dark gray (red) arrows in this sketch. The dashed line shows an additional reversion of the Bloch vector's $x$-$y$ component, implied by Eq. \eqref{eq:spin-expectation}.
}%
\end{figure}

\begin{figure}%
  \includegraphics[width=0.45\textwidth]{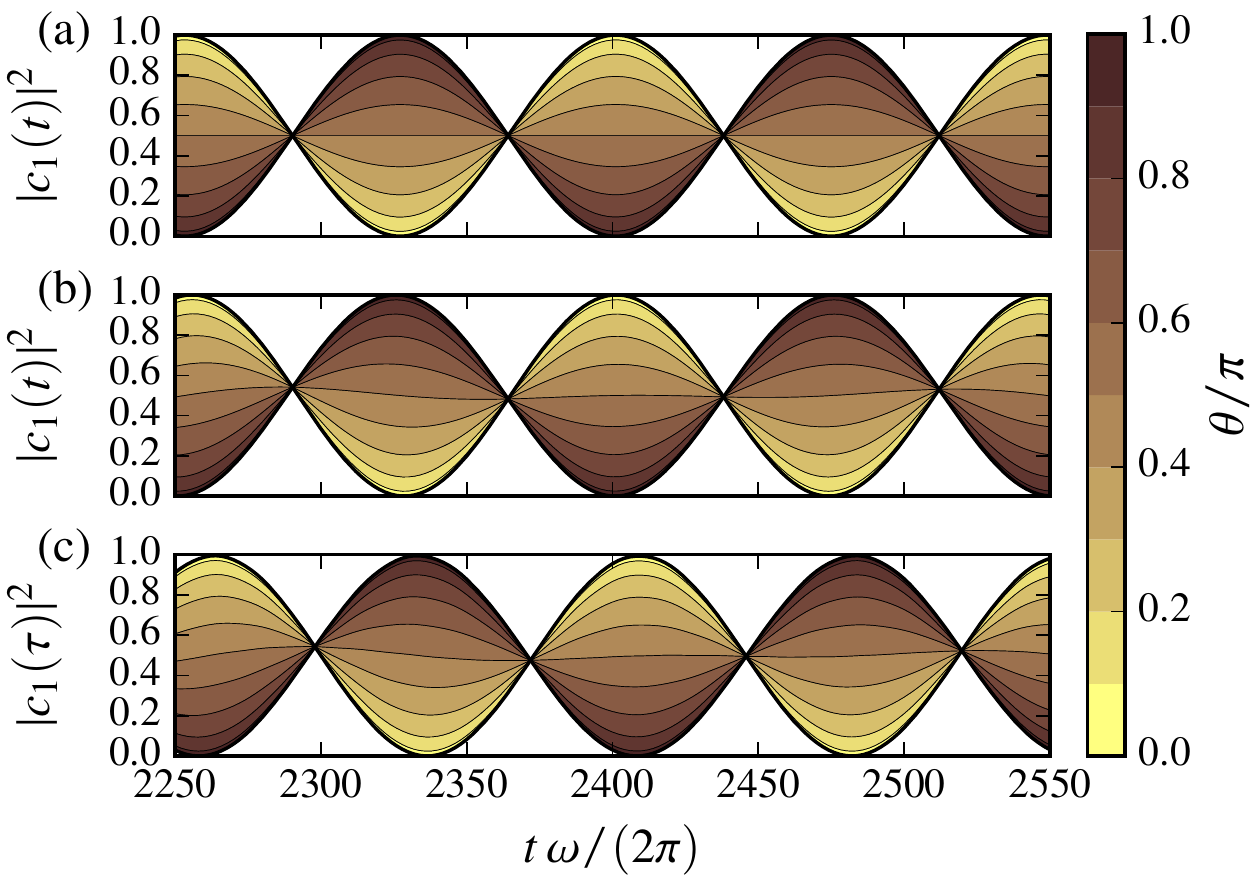}%
  \caption{\label{fig:Bloch_probability}%
(Color online) Spin-dependent diffraction probability. The diffraction probability \eqref{eq:general_diffraction_probability} of finding an electron with initial state \eqref{eq:bloch_state} at final momentum $\hbar k$ is plotted over time. The initial quantum state is evolved in time by the approximation \eqref{eq:spin-rabi-propagator-submatrices} in subplot (a), by the more accurate approximation \eqref{eq:better_propagator_submatricies} in subplot (b) and by a numerical solution according to the differential equations \eqref{eq:dirac_equation_momentum_space} in subplot (c). One can see that the diffraction probability depends on the initial spin configuration of the electron, where $\theta=0$ corresponds to spin-up and $\theta=\pi$ corresponds to spin-down.
}%
\end{figure}

For illustration of the spin properties of the spin propagation matrices $T(t)$ and $R(t)$ in Eq. \eqref{eq:spin-rabi-propagator-submatrices} we assume the quantum state
\begin{equation}
 c_{-1}(0) =
 \begin{pmatrix}
  \cos \frac{\theta}{2} \\ \sin \frac{\theta}{2} e^{i \varphi}
 \end{pmatrix}\,, \qquad
 c_{+1}(0)=0\label{eq:bloch_state}
\end{equation}
for the initial configuration of the electron. This corresponds to an electron which moves with momentum $-\hbar k$ in $z$-direction. The Bloch state parameterization of $c_{-1}(0)$ corresponds to the Bloch vector as sketched in Fig. \ref{fig:bloch_vector}. Inserting the initial state and the matrices \eqref{eq:spin-rabi-propagator-submatrices} into the time-evolution \eqref{eq:time-evolution-notion} results in the final quantum states
\begin{subequations}%
\begin{equation}%
  c_{+1}(t) =
 \begin{pmatrix}
  -i \cos \eta \cos \frac{\theta}{2} \\ -i \sin \eta \sin \frac{\theta}{2} e^{i \varphi}
 \end{pmatrix}\label{eq:diffracted_electron_quantum_state}
\end{equation}%
for the diffracted electron wavefunction and%
\begin{equation}%
  c_{-1}(t) =
 \begin{pmatrix}
  - \sin \eta \cos \frac{\theta}{2} \\ \cos \eta \sin \frac{\theta}{2} e^{i \varphi}
 \end{pmatrix}
\end{equation}\label{eq:evolved_approximated_quantumstates}%
\end{subequations}%
for the undiffracted electron wavefunction. Consequently, one can compute the probabilities
\begin{equation}
 |c_{\pm1}(t)|^2 = \frac{1}{2} \left( 1 \pm \cos \theta \cos 2 \eta \right)\,,\label{eq:diffraction-probability}
\end{equation}
of finding the electron moving in positive or negative $z$-direction.
%
%
Eq. \eqref{eq:diffraction-probability} implies a time-dependence of the diffraction probability by the time-parameter $\eta$ and also a dependence on the original azimutal spin-orientation $\theta$. We are plotting the diffraction probability $|c_{+1}(t)|^2$ in Fig.  \ref{fig:Bloch_probability}(a). The chosen period of time from $2250\, \omega/(2 \pi)$ till $2550\, \omega/(2 \pi)$ corresponds to the one Rabi period $2 \pi/\Omega_R$, starting and ending approximately at the parameter values $\eta \approx 7.5$ and $\eta \approx 8.5$, respectively. In Fig. \ref{fig:Bloch_probability}(b) we also plot the diffraction probability of the time-propagation \eqref{eq:better_propagator_submatricies} of the quantum state \eqref{eq:bloch_state} for comparison. Eq. \eqref{eq:better_propagator_submatricies} is the most accurate available analytic approximation \cite{erhard_bauke_2015_spin}. Furthermore, we plot the numerically simulated time-evolution of the quantum state \eqref{eq:bloch_state} by using the relativistic equations of motion \eqref{eq:dirac_equation_momentum_space} and the commonly used envelope function \cite{ahrens_bauke_2012_spin-kde,ahrens_bauke_2013_relativistic_KDE,bauke_ahrens_2014_spin_precession_1,bauke_ahrens_2014_spin_precession_2,dellweg_mueller_2015_bichromatic_KDE,dellweg_mueller_2015_KDE_pulse_shape,erhard_bauke_2015_spin}
\begin{equation}
 w(t) =
 \begin{cases}
  \sin^2 \frac{\pi t}{2 \,\delta \tau} & \textrm{if } 0 \le t \le \delta \tau \\
  1 & \textrm{if } \delta \tau \le t \le \tau - \delta \tau\\
  \sin^2 \frac{\pi (\tau-t)}{2 \,\delta \tau} & \textrm{if } \tau - \delta \tau \le \tau \,,
 \end{cases}\label{eq:beam-envelope_plateau}
\end{equation}
in Fig. \ref{fig:Bloch_probability}(c). Different simulations with different parameters $\tau$ are carried out for each time $t$ in Fig. \ref{fig:Bloch_probability}(c) and the simulation results $|c_n(\tau)|^2$ at the end of every simulation are plotted. For all simulations we set $\delta \tau=10 \pi/\omega$.

The nice agreement of Fig. \ref{fig:Bloch_probability}(a) with Fig. \ref{fig:Bloch_probability}(b) and Fig. \ref{fig:Bloch_probability}(c) indicates, that the analytic solution \eqref{eq:propagator_submatricies} describes the spin-dynamics well in the chosen period of time. One can see a small retardation of the numeric solution in Fig. \ref{fig:Bloch_probability}(c), as compared to the analytic solutions in Fig. \ref{fig:Bloch_probability}(a) and Fig. \ref{fig:Bloch_probability}(b) of 7.7 laser cycles. This retardation can be explained by the switch on and switch off process of the external field, as discussed in appendix \ref{sec:field_amplitude}. In fact, if one inserts $\delta \tau=10 \pi/\omega$ in Eq. \eqref{eq:sin-plateau_retardation} one obtains a retardation of the scaled time by 6.3 laser cycles.

The quantum dynamics shown here differs fundamentally from previous investigations \cite{ahrens_bauke_2012_spin-kde,ahrens_bauke_2013_relativistic_KDE,bauke_ahrens_2014_spin_precession_1,bauke_ahrens_2014_spin_precession_2,dellweg_awwad_mueller_2016_spin-dynamics_bichromatic_laser_fields} of the Kapitza-Dirac effect in which the propagation of the electron spin could be described by the matrix $U_s(t)$ in Eq. \eqref{eq:su2-representation} with a real-valued unit vector $\vec n$. For example, if the vector $\vec n$ in the set of the approximation parameters \eqref{eq:su-rotation_adaption_first_line2} and \eqref{eq:shifted_xi_parameter2} had the value $(0,0,-1)^T$, the final quantum state of the diffracted electron would be
\begin{equation}
 c_{+1}(t) = -\frac{i}{\sqrt{2}}
 \begin{pmatrix}
  \cos \frac{\theta}{2} e^{i \frac{\xi}{2}} \\ \sin \frac{\theta}{2} e^{i \varphi - i \frac{\xi}{2}}
 \end{pmatrix}\label{eq:rotating_quantum_state}
\end{equation}
instead of \eqref{eq:diffracted_electron_quantum_state}. Then, the diffraction probability would have the time-independent and also spin in-dependent value $|c_{+1}(t)|^2=1/2$. In contrast to that, the diffraction probability is time- and spin-dependent in Fig. \ref{fig:Bloch_probability}. This property can be used to separate spin-up from spin-down electrons in form of a spin filter.

\subsection{Polarization of the electron spin\label{sec:electron_spin-polarization}}

\begin{figure}%
  \includegraphics[width=0.45\textwidth]{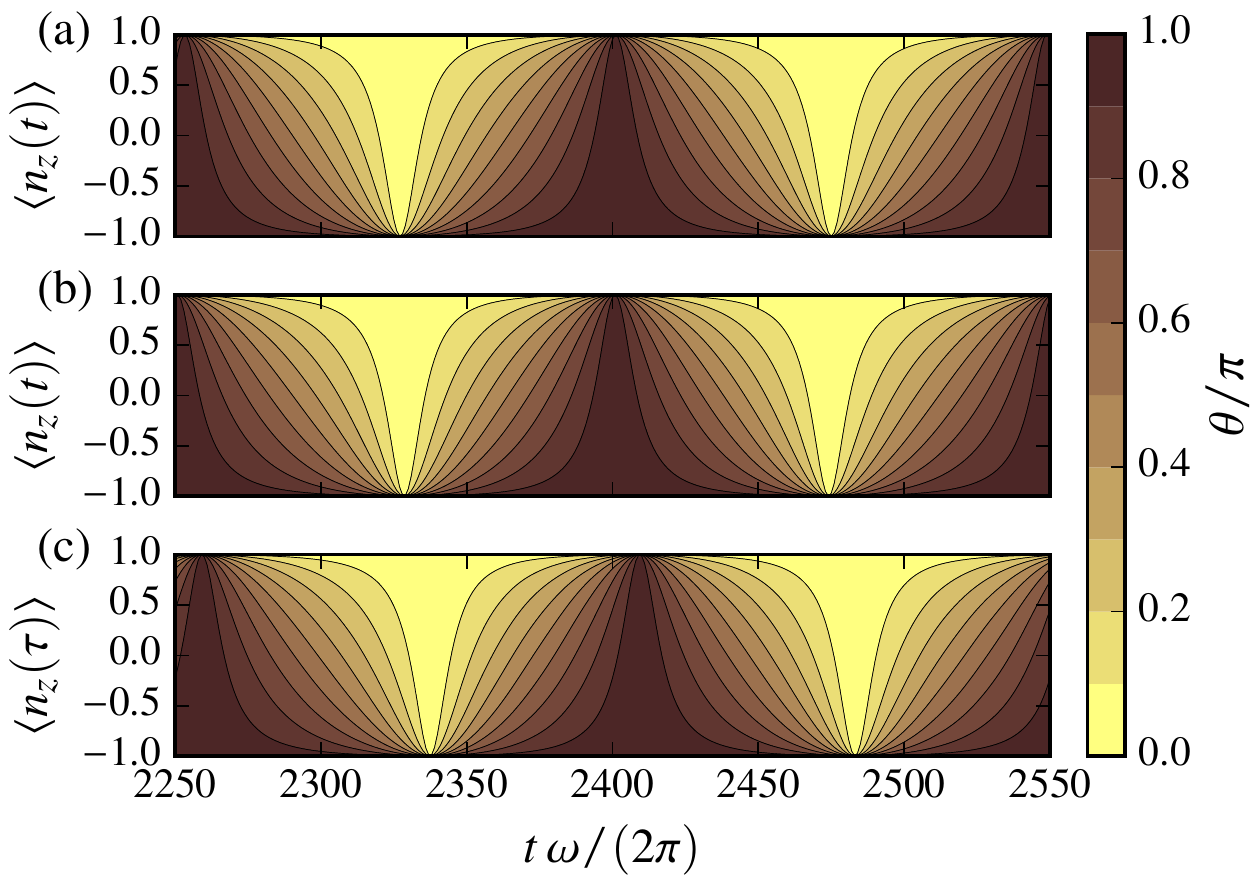}%
  \caption{\label{fig:Bloch_z_component}%
(Color online) Polarization of the electron spin. The $z$-component of the Bloch vector \eqref{eq:bloch_vector} is plotted over time. Similarly as for Fig. \ref{fig:Bloch_probability}, the initial quantum state \eqref{eq:bloch_state} is evolved in time by the approximation \eqref{eq:spin-rabi-propagator-submatrices} in subplot (a), by the more accurate approximation \eqref{eq:better_propagator_submatricies} in subplot (b) and by a numerical solution according to the differential equations \eqref{eq:dirac_equation_momentum_space} in subplot (c). One can observe, that $\langle \vec n(t) \rangle$ is pointing upwards and downwards periodically, implying that the electron spin can be polarized.
}%
\end{figure}

The expectation value of the final quantum state \eqref{eq:diffracted_electron_quantum_state} with respect to the Pauli spin matrices is
\begin{subequations}%
\begin{align}%
 c_{+1}(t)^\dagger \sigma_x c_{+1}(t) &= \frac{1}{2} \sin \theta \sin 2 \eta \cos \varphi \\
 c_{+1}(t)^\dagger \sigma_y c_{+1}(t) &= \frac{1}{2} \sin \theta \sin 2 \eta \sin \varphi \\
 c_{+1}(t)^\dagger \sigma_z c_{+1}(t) &= \frac{1}{2} \left(\cos \theta + \cos 2 \eta \right)\,.\label{eq:spin-expectation_z}
\end{align}\label{eq:spin-expectation}%
\end{subequations}%
This can be normalized by the diffraction probability $|c_{+1}(t)|^2$ of Eq. \eqref{eq:diffraction-probability}
and results in the Bloch vector
\begin{equation}
 \langle \vec n(t) \rangle= \frac{c_{+1}(t)^\dagger \boldsymbol{\sigma} c_{+1}(t)}{|c_{+1}(t)|^2} \label{eq:bloch_vector}
\end{equation}
of the electron spin of the diffracted electron. The $z$-component of the Bloch vector \eqref{eq:bloch_vector} is plotted in Fig. \ref{fig:Bloch_z_component}(a). We also plot the $z$-component of the Bloch vector resulting from the time-propagation \eqref{eq:better_propagator_submatricies} of the initial quantum state \eqref{eq:bloch_state} in Fig. \ref{fig:Bloch_z_component}(b), which is similar to Fig. \ref{fig:Bloch_probability}(b). Also the $z$-component of the Bloch vector from the numerical propagation of the relativistic equations of motion \eqref{eq:dirac_equation_momentum_space} is plotted in Fig. \ref{fig:Bloch_z_component}(c), according to the same procedure as for Fig. \ref{fig:Bloch_probability}(c).

The subplots in Fig. \ref{fig:Bloch_z_component} agree similarly as described for the subplots in Fig. \ref{fig:Bloch_probability}.
This again emphasizes that Eq. \eqref{eq:propagator_submatricies} is a suitable solution of the quantum dynamics in the chosen time period and also confirms the identified retardation \eqref{eq:sin-plateau_retardation} due to the switch on and switch off process of the external field.

In analogy to the diffraction probability, the shown dynamics of the spin-direction of the electron is fundamentally different from dynamics which is related to a real-valued unit vector $\vec n$ in Eq. \eqref{eq:su2-representation}. For example the Bloch vector
\begin{subequations}%
\begin{align}%
 c_{+1}(t)^\dagger \sigma_x c_{+1}(t) &= - \frac{1}{2} \sin \theta \sin (\varphi - 2 \eta) \\
 c_{+1}(t)^\dagger \sigma_y c_{+1}(t) &= \phantom{\pm} \frac{1}{2} \sin \theta \cos (\varphi - 2 \eta) \\
 c_{+1}(t)^\dagger \sigma_z c_{+1}(t) &= \phantom{\pm} \frac{1}{2} \cos \theta \,,
\end{align}%
\end{subequations}%
of the assumed quantum state \eqref{eq:rotating_quantum_state} is rotating with angular velocity $-2 \Omega_R$ around the $z$-axis. In contrast to that, the $z$-component of the electron's spin vector is periodically tilted upwards and downwards in Fig. \ref{fig:Bloch_z_component}, as sketched by the red arrows in Fig. \ref{fig:bloch_vector}. At certain times, for example at $t\approx 2400\cdot2\pi/\omega$ or $t\approx 2480\cdot2\pi/\omega$ the electron is always pointing upwards or downwards respectively, independent of its initial polarization in the $z$-direction. Therefore, it is possible to polarize an initially unpolarized electron spin.

Still, the vector of spin expectation values \eqref{eq:spin-expectation} is flipping its direction in the $x$-$y$-plane with period $\pi/\Omega_R$. This flipping is sketched for illustration as dashed line in Fig. \ref{fig:bloch_vector}. Hence, the spin-flipping dynamics in the $x$-$y$-plane goes along with spin-polarizing and spin-filtering effects along the $z$-direction, in accordance with dynamics reported by Erhard and Bauke \cite{erhard_bauke_2015_spin}, if one accounts for the choice of laser geometry.

\subsection{Distinct spin separation\label{eq:full_spin_filter}}

For $\eta=\pi n,\ n\in\mathbb{N}$ the approximate solution \eqref{eq:spin-rabi-propagator-submatrices} implies that spin-up electrons will be diffracted with momentum reversal in $z$-direction with probability 1 and, likewise, spin-down electrons will remain in their motional state with probability 1. The reversed property, ie. diffraction of spin-down electrons with probability 1 and no diffraction of spin-up electrons with probability 1, is reached for $\eta=\pi/2 + \pi n,\ n\in\mathbb{N}$.

Fig. \ref{fig:Bloch_probability} and Fig. \ref{fig:Bloch_z_component} suggest that Eq. \eqref{eq:spin-rabi-propagator-submatrices} is a good approximation for the value $\eta=8\cdot 2\pi$. This is indeed the case, because one can compute
\begin{equation}
 t^{\prime\prime}=\frac{16\pi-\varphi_0}{\Omega_R}\approx-0.18 \cdot \frac{2 \pi}{\Omega_R}\,,
\end{equation}
for $\eta=8 \cdot 2 \pi$. Then, the necessary condition $|t^{\prime\prime}| \ll 2 \pi/\Omega_S$, which is the requirement that Eq. \eqref{eq:spin-rabi-propagator-submatrices} is a good approximation, is fulfilled well. In the case of $\eta=8 \cdot 2 \pi$ one obtains
\begin{equation}
 T(t) =
 \begin{pmatrix}
  0 & 0 \\
  0 & 1
 \end{pmatrix}\ ,\qquad
 R(t) =
 \begin{pmatrix}
  -i & 0 \\
   0 & 0
 \end{pmatrix}\,,\label{eq:specific_R_T_matrices}
\end{equation}
for the matrices \eqref{eq:spin-rabi-propagator-submatrices}. The action of $T(t)$ and $R(t)$ at the initial conditions in Fig. \ref{fig:simulation} is resulting in the corresponding final configurations in all four subfigures of Fig. \ref{fig:simulation}, as illustrated in Fig. \ref{fig:experimental_setup}. In other words Eq. \eqref{eq:specific_R_T_matrices} displays the propagation matrix of the quantum dynamics in Fig. \ref{fig:simulation} and its illustration in Fig. \ref{fig:experimental_setup}. 

Note, that the time $t$ in Eq. \eqref{eq:specific_R_T_matrices} is related to $\eta$ by Eq. \eqref{eq:shifted_time_quarter_plus} and Eq. \eqref{eq:def_eta} and evaluates to
\begin{equation}
 t = \frac{\eta}{\Omega_R} + \frac{\pi}{4 \Omega_R} = 2401 \frac{\omega}{2 \pi} = 1.27\,\textrm{fs}\,.\label{eq:time_at_eta_8}
\end{equation}
This differs from the total interaction time $\tau$ of the dynamics in Fig. \ref{fig:simulation} which is 6399 laser cycles, corresponding to 3.39\,fs. The reason is the usage of a plateau shaped envelope function \eqref{eq:beam-envelope_plateau} for the simulation in Fig. \ref{fig:Bloch_probability}(c) and Fig. \ref{fig:Bloch_z_component}(c) as compared to the Gaussian shaped envelope function \eqref{eq:beam-envelope} in Fig. \ref{fig:simulation}. It is possible to associate the different field envelopes by a substitution technique which is discussed in appendix \ref{sec:field_amplitude}.
We obtain $\tilde t(\tau)=2400\,\omega/(2 \pi)$ from Eq. \eqref{eq:scaled_total_time} which is fitting to the time in Eq. \eqref{eq:time_at_eta_8}. This demonstrates that the considerations in appendix \ref{sec:field_amplitude} are suitable for relating the quantum dynamics of different time-dependent field amplitudes to each other.

The matrices $T(t)$ and $R(t)$ in Eq. \eqref{eq:specific_R_T_matrices} can be expressed in terms of $U_s(t)$ in Eq. \eqref{eq:su2-representation} by the parameters \eqref{eq:su-rotation_adaption_first_line} with $\xi=\pi/2$ and \eqref{eq:su-rotation_adaption_first_line2} with $\xi=-\pi/2$, respectively. For any set of parameters, the matrices in Eq. \eqref{eq:specific_R_T_matrices} could only be approximated by $U_s(t)$ if an imaginary valued unit vector $\vec n$ was used. In contrast to that, the spin-dynamics as described in the references \cite{ahrens_bauke_2012_spin-kde,ahrens_bauke_2013_relativistic_KDE,bauke_ahrens_2014_spin_precession_1,bauke_ahrens_2014_spin_precession_2,dellweg_awwad_mueller_2016_spin-dynamics_bichromatic_laser_fields} would require a real-valued unit vector $\vec n$ for the parameterization in terms of $U_s(t)$. The real-valued unit vector $\vec n$ implies that the absolute values of all eigenvalues of $U_s(t)$ have the same value $\sqrt{P}$. For the imaginary unit vector $\vec n$ in the representation of $U_s(t)$ the absolute values of the eigenvalues are $0$ and $1$, as for the matrices $T(t)$ and $R(t)$ in Eq. \eqref{eq:specific_R_T_matrices}. In other words $T(t)$ and $R(t)$ are projection matrices times a complex phase. This property differs fundamentally from the unitary propagation of the electron spin in previous descriptions of spin-dynamics in the Kapitza-Dirac effect.

We point out that it is possible to identify similar properties for the quantum state propagation of an electron in a phase-grating \cite{McGregor_Batelaan_2011_TSGM}. Also, spin-dependent electron scattering in a recently proposed, bi-chromatic, interferometric beam splitter \cite{dellweg_mueller_2016_interferometric_spin-polarizer} can be described in terms of a non-unitary spin propagation, as well as spin-dependent electron diffraction in a Kapitza-Dirac effect with three interacting photons of arbitrary polarization \cite{dellweg_mueller_extended_KDE_calculations}, as we have become aware at the final stage of our research.

\section{Conclusion\label{sec:conclusions}}

The effect, which is described in this article allows for the polarization and spin-detection of free electrons due to interaction with a standing, circularly polarized light wave by effectively exchanging two photons only. Our results are presented in the context of high intensity X-ray laser beams of novel facilities, for which the feasibility of electron diffraction with spin effects at similar parameters has been discussed already in an earlier investigation \cite{ahrens_bauke_2012_spin-kde}. More details on the experimental feasibility are also considered in \cite{dellweg_mueller_2016_interferometric_spin-polarizer}. Furthermore, the experimental community discusses the implementation of spin-dependent electron scattering with light in the optical regime \cite{McGregor_Batelaan_2015_two_color_spin} and a perturbative variant is possible in terms of higher order Compton scattering in the high energy regime \cite{Ivanov_2004_polarization_effects_photon_electron,Boca_2012_spin_effects_nonlinear_compton_scattering,Krajewska_2014_spin_polarization_nonlinear_compton_and_thomson_scattering}. We point out that the effect takes place within a resonance peak of the diffraction process (see \cite{ahrens_bauke_2013_relativistic_KDE} and \cite{dellweg_awwad_mueller_2016_spin-dynamics_bichromatic_laser_fields}). A laser frequency uncertainty of $2.3\cdot 10^{18}\,$Hz would be located inside the half width of this resonance peak of the considered two-photon interaction. This also implies that a momentum uncertainty of the electron's momentum component in laser propagation direction has to be below $1.5\,\textrm{keV}/c$, which can be concluded from the requirement of energy and momentum conservation (see discussion in section 8.5 and 2.2.4 in reference \cite{ahrens_2012_phdthesis_KDE}). On the other hand, simulations similar to Fig. 4 (c) show that non-zero electron momenta perpendicular to the laser propagation direction have almost no influence on the diffraction probability as long as this transverse momentum component is smaller than $0.1 mc \approx 51\,$keV. We also mention that the assumed external potential of the laser field \eqref{eq:vector_potential} is not accounting for a spacial envelope in the context of the plane-wave ansatz in this work. The effect of a space-dependent pulse envelope on the spin-dependent diffraction dynamics is a remaining aspect, which should be studied in the future.

Our work has shown that the quantum dynamics is already described properly by the Pauli equation with relativistic corrections \eqref{eq:relativistic_pauli_equation}, which is consistent with the dynamics from the Dirac equation \eqref{eq:dirac-equation}. Nevertheless, the effect of spin-dependent diffraction only occurs for weakly relativistic parameters of the light wave's frequency and its field amplitude. The electron in an external field is treated in terms of the most fundamental description in particle physics (Dirac equation in external fields) as compared to effective theories for example in solid state physics or quantum optics. Therefore, the effect of spin-dependent diffraction could be a test bed for examining relativistic quantum dynamics at the fundamental level, if the required external fields can be provided accurately in experiment. At the current stage we don't expect significant sensitivity on new physics from the effect, unless drastic changes to the standard model would be applied. However, further studies would be needed to make authentic statements on fundamental effects beyond the standard model in particle physics.

Since the effect is also sensitive on the pulse amplitude and pulse duration of the laser field it could be useful for beam diagnosis. Finally, we point out that the propagation of the quantum state is described by a unitary transformation. Therefore, the effect can be reversed, provided that a high fidelity experimental setup is available.

\subsection*{Acknowledgements}
S.A. thanks for help and feedback from Shi-Yao Zhu, Chang-Pu Sun, Tilen Cadez, Heiko Bauke and Carsten M\"uller. This work has been supported by the National Basic Research Program of China (Grant No. 2016YFA0301201 and No. 2014CB921403), by the NSFC (Grant No. 11650110442 and No. 11421063 and No. 11534002) and the NSAF (Grant No. U1530401).

\appendix
\section{The effect of the time-dependent field amplitude\label{sec:field_amplitude}}

Assume the constant field amplitude $A$ would be replaced by the time-dependent amplitude $A \,w(t)$ in Eq. \eqref{eq:simplified_pauli_equation}, with corresponding time-dependent frequencies
\begin{equation}
 \Omega_R(t) = \Omega_R \, w(t)^2 \quad \textrm{and} \quad \Omega_S(t) = \Omega_S \, w(t)^2\,.
\end{equation}
Then $w(t)$ is just appearing as time-dependent prefactor of the coefficients of the differential equation
\begin{equation}
 i \dot c_{\pm 1}(t) = w(t)^2 (\Omega_R \mathbf{1} + \Omega_S \sigma_z) c_{\mp 1}(t)\,.\label{eq:time-dependent_pauli_equation}
\end{equation}
We want to transform this equation to a formally equivalent version of Eq. \eqref{eq:simplified_pauli_equation} with reparameterized coefficients $\tilde c\left(\tilde t\right)$ of a scaled time
\begin{equation}
 \tilde t(t) = \int_0^t w(s)^2 ds \,,
\end{equation}
such that $c(t)=\tilde c\left(\tilde t(t)\right)$. Such a scaled time corresponds to the warped time parameter \cite{ahrens_2012_phdthesis_KDE} or the the action parameter \cite{dellweg_mueller_2015_bichromatic_KDE}. The new coefficients imply
\begin{multline}
 \dot c_n(t) = \frac{\partial}{\partial t} c_n(t)= \frac{\partial}{\partial t} \tilde c_n\left(\tilde t(t)\right) \\
 = \frac{\partial\tilde t(t)}{\partial t} \frac{\partial}{\partial \tilde t} \tilde c_n\left(\tilde t\right)=w(t)^2 \frac{\partial}{\partial \tilde t} \tilde c_n\left(\tilde t\right)
\end{multline}
due to the inner derivative. Plugging this back into \eqref{eq:time-dependent_pauli_equation} results in
\begin{equation}
 i \frac{\partial}{\partial \tilde t}\tilde c\left(\tilde t\right) = (\Omega_R \mathbf{1} + \Omega_S \sigma_z) \tilde c_{\mp 1}\left(\tilde t\right)\label{eq:new_time-dependent_pauli_equation}
\end{equation}
which is formally equivalent to Eq. \eqref{eq:simplified_pauli_equation}, as desired. Integrating the scaled time for the whole interaction time $\tau$ yields the time
\begin{equation}
 \tilde t(\tau) = \int_0^{\tau} w(s)^2 ds = \int_0^{\tau} \sin^4 \left(\frac{\pi s}{\tau}\right) ds =\frac{3}{8} \tau
 \,,\label{eq:scaled_total_time}
\end{equation}
for the Gaussian envelope function \eqref{eq:beam-envelope}. For the plateau-shaped envelope function \eqref{eq:beam-envelope_plateau} one obtains
\begin{equation}
 \tilde t(\tau) = \int_0^{\tau} w(s)^2 ds = \tau - \frac{5}{4} \delta \tau\label{eq:sin-plateau_retardation}
\end{equation}
in a similar calculation.
\vspace{0.5 cm}

\section{Comparison with more accurate solution\label{sec:more_acurate_solution}}

We want to compare the approximate solution \eqref{eq:propagator_submatricies} of section \ref{sec:approximative_solution} with the solution given in \cite{erhard_bauke_2015_spin} which also accounts for the electron momenta $3 \hbar$ and $-3 \hbar k$. By performing an analog derivation, the differential equation \eqref{eq:momentum_space_pauli_equation} can first be written in matrix notion as
\begin{equation}
 i
 \begin{pmatrix}
  \dot c_{-3} \\ \dot c_{-1} \\ \dot c_{+1} \\ \dot c_{+3}
 \end{pmatrix}
 =
 \begin{pmatrix}
  9 \Omega_k \mathbf{1} & M & 0 & 0 \\
  M & \Omega_k \mathbf{1} & M & 0 \\
  0 & M & \Omega_k \mathbf{1} & M \\
  0 & 0 & M & 9 \Omega_k \mathbf{1}
 \end{pmatrix}
 \begin{pmatrix}
  c_{-3} \\ c_{-1} \\ c_{+1} \\ c_{+3}
 \end{pmatrix}\label{eq:approximated_quantum_system}
\end{equation}
with the spin-dependent coupling matrix
\begin{equation}
 M = \Omega_R \mathbf{1} + \Omega_S \sigma_z = 
  \begin{pmatrix}
  \Omega_R + \Omega_S & 0 \\
  0 & \Omega_R - \Omega_S
 \end{pmatrix}\,.
\end{equation}
A constant from the ponderomotive potential, which causes a global phase with oscillation frequency $2 \Omega_R$ can be omitted by choice of a suitable gauge. Similarly as in reference \cite{erhard_bauke_2015_spin}, we use a computer algebra system and a simplification for expressions of the form
\begin{multline}
 \sqrt{\left[8 \Omega_k - (\Omega_R + \Omega_S)\right]^2 + 4 (\Omega_R + \Omega_S)^2} \\
 = 8 \Omega_k - \Omega_R - \Omega_S + \frac{(\Omega_R + \Omega_S)^2}{4 \Omega_k} +\dots\,,\label{eq:sqrt_approximation}
\end{multline}
for the case of small frequencies $\Omega_R\ll\Omega_k$ and $\Omega_S\ll\Omega_k$ and therewith small numbers $\Omega_R/\Omega_k$ and $\Omega_S/\Omega_k$. For the matrix in Eq. \eqref{eq:approximated_quantum_system}, we obtain the approximated eigenenergies
\begin{subequations}
\begin{align}
 \epsilon_1 &\approx \epsilon_0 + \Omega_R + \Omega_S - \Delta \\
 \epsilon_2 &\approx \epsilon_0 - \Omega_R + \Omega_S + \Delta \\
 \epsilon_3 &\approx \epsilon_0 + \Omega_R - \Omega_S + \Delta \\
 \epsilon_4 &\approx \epsilon_0 - \Omega_R - \Omega_S - \Delta \\
 \epsilon_5 &\approx \epsilon_0 + 8 \Omega_k - \Delta \\
 \epsilon_6 &\approx \epsilon_0 + 8 \Omega_k + \Delta \\
 \epsilon_7 &\approx \epsilon_0 + 8 \Omega_k + \Delta \\
 \epsilon_8 &\approx \epsilon_0 + 8 \Omega_k - \Delta \,,
\end{align}\label{eq:eigenvalues_rewritten}%
\end{subequations}
where we have introduced the frequency of an energy shift
\begin{equation}
 \epsilon_0 = \Omega_k - \frac{\Omega_R^2 + \Omega_S^2}{8 \Omega_k}
\end{equation}
and the frequency of higher order corrections of the quantum dynamics
\begin{equation}
 \Delta = \frac{\Omega_R \Omega_S}{4 \Omega_k}\,.
\end{equation}
The frequency $\epsilon_0$ will be omitted in the following calculation, as it causes an additional, time-dependent phase of the quantum system \eqref{eq:approximated_quantum_system} which can be removed by choice of a suitable gauge. The corresponding approximated eigenvectors of the eigenvalues \eqref{eq:eigenvalues_rewritten} are
\pagebreak
\begin{widetext}
\begin{subequations}
\begin{equation}
 \vec v_1=\left( 1, 0, - \frac{8 \Omega_k}{\Omega_R + \Omega_S} + 1 - \frac{\Omega_R + \Omega_S}{8 \Omega_k} , 0, - \frac{8 \Omega_k}{\Omega_R + \Omega_S} + 1 - \frac{\Omega_R + \Omega_S}{8 \Omega_k} , 0 , 1, 0 \right)^T
\end{equation}
\begin{equation}
 \vec v_2=\left( 0, -1, 0, \frac{8 \Omega_k}{\Omega_R - \Omega_S} + 1 - \frac{-\Omega_R + \Omega_S}{8 \Omega_k} , 0, - \frac{8 \Omega_k}{\Omega_R - \Omega_S} - 1 + \frac{-\Omega_R + \Omega_S}{8 \Omega_k} , 0 , 1 \right)^T
\end{equation}
\begin{equation}
 \vec v_3=\left( 0, 1, 0, - \frac{8 \Omega_k}{\Omega_R - \Omega_S} + 1 + \frac{-\Omega_R + \Omega_S}{8 \Omega_k} , 0, - \frac{8 \Omega_k}{\Omega_R - \Omega_S} + 1 + \frac{-\Omega_R + \Omega_S}{8 \Omega_k} , 0 , 1 \right)^T
\end{equation}
\begin{equation}
 \vec v_4=\left( -1, 0, \frac{8 \Omega_k}{\Omega_R + \Omega_S} + 1 + \frac{\Omega_R + \Omega_S}{8 \Omega_k} , 0, - \frac{8 \Omega_k}{\Omega_R + \Omega_S} - 1 - \frac{\Omega_R + \Omega_S}{8 \Omega_k} , 0 , 1, 0 \right)^T
\end{equation}
\begin{equation}
 \vec v_5=\left( 1, 0, \frac{\Omega_R + \Omega_S}{8 \Omega_k} , 0, \frac{\Omega_R + \Omega_S}{8 \Omega_k} , 0 , 1, 0 \right)^T
\end{equation}
\begin{equation}
 \vec v_6=\left( 0 , -1, 0, \frac{-\Omega_R + \Omega_S}{8 \Omega_k} , 0, -\frac{-\Omega_R + \Omega_S}{8 \Omega_k} , 0 , 1\right)^T
\end{equation}
\begin{equation}
 \vec v_7=\left( 0 , 1, 0, \frac{\Omega_R - \Omega_S}{8 \Omega_k} , 0, \frac{\Omega_R - \Omega_S}{8 \Omega_k} , 0 , 1\right)^T
\end{equation}
\begin{equation}
 \vec v_8=\left(-1, 0, \frac{\Omega_R + \Omega_S}{8 \Omega_k} , 0, -\frac{\Omega_R + \Omega_S}{8 \Omega_k} , 0 , 1, 0 \right)^T\,.
\end{equation}\label{eq:eigenvectors}%
\end{subequations}
\end{widetext}
For the limit $\Omega_k \gg \Omega_R > \Omega_S$ as used in \cite{erhard_bauke_2015_spin}, the eigenvectors are approximated by
\begin{multline}
\left(\vec v_1,\vec v_2,\vec v_3,\vec v_4,\vec v_5,\vec v_6,\vec v_7,\vec v_8\right) \\
= \frac{1}{\sqrt{2}}
\begin{pmatrix}
  0 &  0 &  0 &  0 & 1 &  0 & 0 &  1 \\
  0 &  0 &  0 &  0 & 0 &  1 & 1 &  0 \\
  1 &  0 &  0 &  1 & 0 &  0 & 0 &  0 \\
  0 &  1 &  1 &  0 & 0 &  0 & 0 &  0 \\
  1 &  0 &  0 & -1 & 0 &  0 & 0 &  0 \\
  0 & -1 &  1 &  0 & 0 &  0 & 0 &  0 \\
  0 &  0 &  0 &  0 & 1 &  0 & 0 & -1 \\
  0 &  0 &  0 &  0 & 0 & -1 & 1 &  0
\end{pmatrix}\,,\label{eq:approximated_eigenvectors}
\end{multline}
and normalized to $1$ here. We are interested in the time-evolution of the quantum states $c_{+1}(t)$ and $c_{-1}(t)$ and point out that the first 4 approximated eigenvectors in Eq. \eqref{eq:approximated_eigenvectors} form a closed subspace of these states. Within this subspace, the time-evolution of the vector of expansion coefficients 
\begin{equation}
 C(t) = \left( c_{-1}^\uparrow(t),c_{-1}^\downarrow(t), c_{+}^\uparrow(t),c_{+1}^\downarrow(t) \right)^T
\end{equation}
has an equivalent expression for the time-evolution \eqref{eq:time-evolution-notion} and can be written as $C(t) = U(t) C(0)$ with the form of the propagator
\begin{equation}
 U(t)=
 \begin{pmatrix}
  T(t) & R(t) \\
  R(t) & T(t)
 \end{pmatrix}\,.\label{eq:propagator}
\end{equation}
The time-evolution can be computed by making use of the matrix exponential
\begin{equation}
 U(t) = V e^{-i D t} V^{-1}\,,
\end{equation}
in which we are using the eigenvector subspace matrix
\begin{equation}
 V={V^{-1}}^\dagger=\frac{1}{\sqrt{2}}
 \begin{pmatrix}
  1 &  0 &  0 &  1 \\
  0 &  1 &  1 &  0 \\
  1 &  0 &  0 & -1 \\
  0 & -1 &  1 &  0
 \end{pmatrix}
\end{equation}
and the corresponding diagonal matrix of eigenvalues $D=\textrm{diag}(\epsilon_1,\epsilon_2,\epsilon_3,\epsilon_4)$.
\noindent From the property $U(t)^\dagger U(t)=\textrm{id}_4$ we note that
\begin{align}
 |T|^2 + |R|^2 = \mathbf{1} \\
 T^\dagger R + R^\dagger T = 0
\end{align}
holds, where $\textrm{id}_4$ is the $4\times4$ identity matrix. Thus $R(t)$ and $T(t)$ are reflection and transmission matrices.

\noindent An explicit expression of $U(t)$ is
\begin{widetext}
\begin{equation}
U(t)=\frac{1}{2}
 \begin{pmatrix}
  e^{-i \epsilon_1 t} + e^{-i \epsilon_4 t} & 0 & e^{-i \epsilon_1 t} - e^{-i \epsilon_4 t} & 0 \\
  0 & e^{-i \epsilon_2 t} + e^{-i \epsilon_3 t} & 0 & - e^{-i \epsilon_2 t} + e^{-i \epsilon_3 t} \\
  e^{-i \epsilon_1 t} - e^{-i \epsilon_4 t} & 0 & e^{-i \epsilon_1 t} + e^{-i \epsilon_4 t} & 0 \\
  0 & -e^{-i \epsilon_2 t} + e^{-i \epsilon_3 t} & 0 & e^{-i \epsilon_2 t} + e^{-i \epsilon_3 t}
 \end{pmatrix}\,,\label{eq:general_time-evolution}
\end{equation}
\end{widetext}
from which one can read off the matrices
\begin{subequations}
\begin{align}
T(t) &= \textrm{diag}(\cos[(\Omega_R + \Omega_S)t] e^{i\Delta t}, \nonumber\\
& \qquad\qquad\qquad\cos[(\Omega_R - \Omega_S)t] e^{-i\Delta t})\,,\label{eq:better_propagator_A}\\
R(t) &= \textrm{diag}(-i\sin[(\Omega_R + \Omega_S)t] e^{i\Delta t} , \nonumber\\
& \qquad\qquad\qquad-i\sin[(\Omega_R - \Omega_S)t] e^{-i\Delta t}) \,.\label{eq:better_propagator_B}
\end{align}\label{eq:better_propagator_submatricies}%
\end{subequations}
Note again, that $\epsilon_0$ has been omitted here.

The frequency $\Delta$ scales with the fourth power of the field amplitude $A$, while $\Omega_R$ and $\Omega_S$ only scale with the square of $A$. Therefore, for small fields the frequency $\Delta$ is smaller than $\Omega_R$ and $\Omega_S$, which is the case for the parameters chosen in Fig. \ref{fig:simulation}. Thus, one may approximate
\begin{equation}
 e^{i\Delta t} \approx 1 + i \Delta t \approx 1
\end{equation}
on time scales which are much shorter than the period $2 \pi/\Delta$. In this case, the matrices $T(t)$ and $R(t)$ in \eqref{eq:better_propagator_submatricies} change into the simpler solution \eqref{eq:propagator_submatricies} of section \ref{sec:approximative_solution}.

\bibliographystyle{apsrev4-1}
\bibliography{bibliography}

\end{document}